\documentstyle[12pt]{article}
\begin{document}
\setlength{\unitlength}{1mm}

\begin{center}
{\Large\bf  On One-Loop Renormalization of Black-Hole }\\
\end{center}
\begin{center}
{\Large\bf Entropy}\\
\end{center}

\bigskip\bigskip

\begin{center}
{\bf Dmitri V. Fursaev}\footnote{e-mail: dfursaev@phys.ualberta.ca}
and {\bf Sergey N. Solodukhin}\footnote{e-mail:
solod@thsun1.jinr.dubna.su}
\end{center}

\begin{center}
{{\it Bogoliubov Laboratory of Theoretical Physics, \\
Joint Institute for Nuclear Research, \\
141 980 Dubna, Moscow Region, Russia}\\}
\end{center}

\bigskip\bigskip\bigskip

\begin{abstract}
One-loop divergences appearing in the entropy of a quantum black hole
are proven
to be completely eliminated by the standard renormalization of both
the gravitational constant and other coefficients by the $R^2$-terms
in
the effective gravitational action. The essential point
of the proof is that
due to the higher order curvature terms the entropy differs from the
Bekenstein-Hawking one in the Einstein gravity by the contributions
depending on the internal and external geometry of the horizon
surface.
\end{abstract}

\vspace{7cm}

{\it PACS number(s): 04.60.+n, 12.25.+e, 97.60.Lf, 11.10.Gh}

\newpage
\baselineskip=.8cm

It is remarkable and a long-known fact that there are analogs
of the thermodynamical laws for
classical black holes
\cite{Bekenstein-Hawking}.
The central point of this analogy is the Bekenstein-Hawking entropy
identified with the area of the black-hole horizon.
However, this entropy so far has not received
statistical interpretation as a quantity counting the number of
states.
Moreover, recent attempts to develop such a statistical approach (for
a
review see \cite{Bekenstein}) discovered divergences
$S_{div}$ in the entropy which are proportional to area of the
horizon surface
\cite{a1}.  The same feature
is inherent in the "entanglement" entropy that appears in ordinary
quantum
theory under tracing out a part of the pure state which resides
inside a
region of space \cite{Srednicki}.  Analogously, one can interpret
$S_{div}$ as divergent part of the entanglement entropy  related with
the loss, for
an external observer, of information about field excitations located
inside the horizon \cite{Sorkin}.  The physical reason for $S_{div}$
to occur in this way is correlations across the
horizon between inside and outside quantum fluctuations.

On the other hand, the infinite value of the entropy is the
consequence of the infinite number
of quantum states with arbitrary large energies which  exist
in  the vicinity of the horizon.
This contradicts to the obvious finiteness of the Bekenstein-Hawking
entropy
appearing from the thermodynamical analogy
and gives rise to the problem: how different definitions of the black
hole
entropy, statistical
and thermodynamical, are related \cite{Frolov}.

A resolution of this contradiction has been proposed in  papers
\cite{SU},
\cite{Jacobson} arguing that for the black hole of a large mass, when
the
space-time can be approximated by the Rindler metric, $S_{div}$
has the same origin as
the conventional ultraviolet divergences that are eliminated by the
standard renormalization of the Newton constant $G$.
In this way, a natural question arises, whether is the resolution
due to
\cite{SU} and \cite{Jacobson} of this puzzle {\it ad hoc},
valid only for a given case, or  there
is a universal statement to be strictly proved for arbitrary black
holes?
Also, is the renormalization of $G$ sufficient to get rid of  all the
divergences?

Some progress has been achieved in the subsequent works.
Thus, it has shortly been realized that for the Schwarzschild black
hole
$S_{div}$ is not reduced only to the horizon area \cite{a3}
and is determined also by the external geometry of the horizon.
In this case,
the addition is removed by the renormalization of a
gravitational coupling at the $R^2$-term necessarily generated in the
effective action
by quantum corrections \cite{a3}.
Then, the complete form of the divergent terms appearing on the
horizon surface has been derived explicitly for an arbitrary
static black-hole geometry in \cite{a4}, which allowed one to find
out all
divergent corrections to the black hole entropy \cite{S}, \cite{F}.

\bigskip\bigskip

Now the aim of our Letter is to give a general proof that
$S_{div}$ treated as  one-loop ultaraviolet divergences can be
completely removed from
the complete black hole entropy
under standard renormalization of the Newton constant $G$
and other couplings in the effective
gravitational action at the second order curvature terms.
In other words, we show that the bare tree-level and
$S_{div}$ pieces of the black hole entropy appear in  a combination
to reproduce tree-level entropy expressed through the renormalized
constants.

\bigskip\bigskip

As in quantum theory
quantum corrections in curved space-time  are known to result in
higher order curvature
contributions to the Einstein action \cite{a16},
we begin our consideration with the action functional
\begin{equation}
W=\int \sqrt{g}d^4x\left(-{1 \over 16\pi G}(R +2\Lambda)+ c_1 R^2 +
c_2R^{\mu\nu}
R_{\mu\nu} + c_3 R^{\mu\nu\lambda\rho}R_{\mu\nu\lambda\rho}\right)~~~
\label{locpart}
\end{equation}
where $G$ and $c_i$ are the bare gravitational couplings and
$\Lambda$ is the cosmological constant.
Due to topological properties,
only two of these couplings are independent in the four-dimensional
theory.

We will follow the Gibbons-Hawking path integral approach to the
gravitational thermodynamics \cite{GH} to give our analysis a
transparent
statistical meaning\footnote{That this approach gives the adequate
description
of the "entanglement" entropy has been demonstrated in \cite{SS}.}.
In this approach, the functional (\ref{locpart}), being
considered on the Euclidean section ${\cal M}_{\beta}$ of the
corresponding space-time with period
$\beta$ in time, is associated, in the semiclassical approximation,
with the tree-level free energy of the system at temperature $T=\beta
^{-1}$
\begin{equation}
F(\beta)=\beta^{-1}W(\beta)~~~.
\label{freeen}
\end{equation}
Nothing unusual happens as compared to thermodynamics in the
Minkowsky space
when space-time possesses a globally defined time-like Killing vector
field that is not null anywhere.
A nontrivial point appears in the presence of the Killing horizon,
as in the case of a black hole geometry.
In this case, for arbitrary
temperature $\beta ^{-1}$
the Euclidean manifold ${\cal M}_{\beta}$ has conical singularities
at
the horizon surface $\Sigma$, in the vicinity of which it
topologically looks
as a space product $C_{\beta}\times\Sigma$ of a two-dimensional cone
$C_{\beta}$
and the horizon surface $\Sigma$.
This leads to a specific Hawking temperature
$\beta^{-1}=\beta_H^{-1}$ for which the Euclidean manifold is
regular.
The black-hole thermodynamics is considered at this temperature.
However,
to get the entropy
using the standard definition
\begin{equation}
S(\beta_H)=\left(\beta{\partial \over \partial \beta}
-1\right)W(\beta)|_{\beta=\beta _H}
\label{entropy}
\end{equation}
we must let $\beta$ be slightly different from $\beta_H$. This
procedure being applied to the action (\ref{locpart}) faces a
difficulty
due to the terms of higher order in curvature which turn out to be
ill-defined on the conical singularities.

There is a method how to avoid this problem \cite{FS} when one
approximates
${\cal M}_{\beta}$  by a sequence of smooth manifolds
${\tilde {\cal M}}_{\beta}$ converging to
${\cal M}_{\beta}$. For the "regularized" spaces ${\tilde {\cal
M}}_{\beta}$
the action (\ref{locpart}) is well-defined and
in the limit ${\tilde {\cal M}}_{\beta} \rightarrow{\cal M}_{\beta}$
we get the following formulas \cite{FS}
\begin{equation}
\int_{{\cal M}_{\beta}}R=\alpha\int_{{\cal M}_{\beta _H}} R
+4\pi(1-\alpha)\int_{\Sigma}~~~,
\label{1}
\end{equation}
\begin{equation}
\int_{{\cal M}_{\beta}}R^2=\alpha\int_{{\cal M}_{\beta _H}} R^2
+8\pi(1-\alpha)\int_{\Sigma}R+O((1-\alpha)^2)~~~,
\label{2}
\end{equation}
\begin{equation}
\int_{{\cal M}_{\beta}}R^{\mu\nu}R_{\mu\nu}=\alpha\int_{{\cal
M}_{\beta _H}}
R^{\mu\nu}R_{\mu\nu}
+4\pi(1-\alpha)\int_{\Sigma}R_{\mu\nu}n_i^{\mu}n_i^{\nu}+O((1-\alpha)^
2)~~~,
\label{3}
\end{equation}
\begin{equation}
\int_{{\cal M}_{\beta}}R^{\mu\nu\lambda\rho}R_{\mu\nu\lambda\rho}
=\alpha\int_{{\cal M}_{\beta _H}}
R^{\mu\nu\lambda\rho}R_{\mu\nu\lambda\rho}
+8\pi(1-\alpha)\int_{\Sigma}R_{\mu\nu\lambda\rho}n^{\mu}_in^{\lambda}_
i
n^{\nu}_j n^{\rho}_j
+O((1-\alpha)^2)~~~,
\label{4}
\end{equation}
where $\alpha=\beta/\beta_H$ and $n^{\mu}_i$ are two orthonormal
vectors
orthogonal to $\Sigma$. The first integrals in the right hand side of
(\ref{1})-(\ref{4})
are defined on the smooth space at $\beta=\beta_H$; they are
proportional to
$\beta$ and do not affect $S(\beta_H)$. As for the
terms $O((1-\alpha)^2)$ in (\ref{2})-({\ref{4}), they depend on
the regularization prescription and turn out to be singular
in the limit ${\tilde {\cal M}}_{\beta} \rightarrow{\cal M}_{\beta}$,
but
they do not contribute to the entropy and energy at the Hawking
temperature
$(\alpha=1)$.
Indeed, from (\ref{1})-(\ref{4}) one obtains for $S$ the following
integral over the horizon surface $\Sigma$:
$$
S(G,c_i)=
\lim_{{\tilde {\cal M}}_{\beta} \rightarrow{\cal M}_{\beta}}
\left(\beta{\partial \over \partial \beta}
-1\right)W({\tilde {\cal M}}_{\beta})(G,c_i,\Lambda)
|_{\beta=\beta _H}
$$
\begin{equation}
={1 \over 4G} A_{\Sigma}-
\int_{\Sigma}\left(8\pi c_1 R + 4\pi c_2 R_{\mu\nu}n^{\mu}_in^{\nu}_i
+ 8\pi c_3
R_{\mu\nu\lambda\rho}n^{\mu}_in^{\lambda}_in^{\nu}_jn^{\rho}_j\right)~
{}~~
\label{locentropy}
\end{equation}
where $A_{\Sigma}$ is the horizon area.
Remarkably, this expression differs from the Bekenstein-Hawking
entropy
$S=A_{\Sigma}/4G$ in
the Einstein gravity by the contributions depending on both internal
and
external geometry of the horizon due to the high curvature
terms in (\ref{locpart})
(The $\Lambda$-term does not
appear in (\ref{locentropy})
explicitly because the volume of ${\cal M}_{\beta}$ is proportional
to $\beta$.). However, it is easy to see that the effect
of internal geometry of $\Sigma$ is reduced to the integral curvature
of this surface which, being a topological invariant, is an
irrelevant
constant addition to the entropy.
It is worth noting that exactly
the same expression can be derived by the Noether charge method
suggested
by Wald \cite{Wald}. The difference between two approaches is that
Wald's
method seems to be more general, but it is defined "on-shell",
whereas the
above derivation of (\ref{locentropy}) did not operate with the
equations of
motion.

Consider now quantum theory on the black-hole background. For a
massive
scalar field the one-loop effective action reads
\begin{equation}
W_{eff}=W+\frac 12 \log\det (-\Box +m^2)~~~.
\label{effact}
\end{equation}
To define this action on the singular manifold ${\cal M}_{\beta}$,
we make use of the same procedure going to ${\tilde {\cal M}}_\beta$.
On the smoothed space $W_{eff}$ consists of the finite $W_{fin}$
and ultraviolet divergent $W_{div}$ parts. The latter has the same
structure as the bare functional (\ref{locpart}). For instance, in
the
dimensional regularization with the parameter $\epsilon$ one can
write
\begin{equation}
W_{div}=-{1 \over 32\pi ^2\epsilon}
\left[{(m^2)^2 \over 2}a_0-m^2 a_1
+a_2\right]~~~,
\label{vol}
\end{equation}
where for regular manifold ${\tilde {\cal M}}_\beta$
and the coefficients $a_i$ standardly  appear in the asymptotic
expansion of the heat kernel of
the operator in (\ref{effact})
\begin{equation}
a_1({\tilde {\cal M}}_\beta)=\frac 16\int_{{\tilde{\cal
M}}_{\beta}}R~~~,
\label{a1}
\end{equation}
\begin{equation}
a_2({\tilde {\cal M}}_\beta)=
\int_{{\tilde {\cal M}}_{\beta}}\left({1 \over 180}
R_{\mu\nu\lambda\rho}
R^{\mu\nu\lambda\rho}-{1 \over 180}R_{\mu\nu}R^{\mu\nu}
+{1 \over 72} R^2\right)~~~,
\label{a2}
\end{equation}
and $a_0=\int_{{\tilde {\cal M}}_\beta}$.
These divergences are
taken off by the usual renormalization of the gravitational
couplings $G$ and $c_i$ of the bare classical action (see
(\ref{locpart}))
\cite{a16}
\begin{equation}
W({\tilde {\cal M}}_{\beta})(G,c_i,\Lambda) +
W_{div}({\tilde {\cal M}}_{\beta})(\epsilon) =
W({\tilde {\cal M}}_{\beta})(G^{ren},c_i^{ren},\Lambda^{ren})
\label{renW}
\end{equation}
where $G^{ren}$ and $c_i^{ren}$ are the renormalized couplings
expressed through
the bare ones and the ultraviolet cut-off parameter $\epsilon$
\begin{equation}
{1 \over G^{ren}}=
{1 \over G}-{m^2 \over 12\pi \epsilon}~~,~~{\Lambda^{ren} \over
G^{ren}}={\Lambda \over G}
+{(m^2)^2 \over 8\pi \epsilon}~~,
{}~~c_1^{ren}=c_1-{1 \over 32 \pi ^2
72 \epsilon}~~~,
\label{const}
\end{equation}
$$
c_{2,ren}=c_2+{1 \over 32\pi ^2 180 \epsilon}~~~,
{}~~~c_{3,ren}=c_3-{1 \over 32\pi ^2 180 \epsilon}~~~.
$$

Now one should go to the limit
${\tilde {\cal M}}_{\beta} \rightarrow{\cal M}_{\beta}$.
In this case, the effective action
$W_{eff}$ acquires additional surface ultraviolet infinite terms
which will then be responsible for divergences in the one-loop
black hole entropy. It is clear that the part of new divergences
proportional
to $(1-\alpha)$, which would come
from
the $W_{div}({\tilde {\cal M}}_\beta)$ according to
(\ref{1})-(\ref{4}),
have already been  removed in due course of the standard
renormalization
(\ref{renW}) of the gravitational couplings and they will not appear
in the entropy.

However, in the limit when the smoothed geometry ${\tilde {\cal M}}
_\beta$ is replaced by the singular one the finite part of the
effective
action $W_{fin}$ might also diverge at the horizon,
as has been demonstrated in the two-dimensional case
where $W_{fin}$ is exactly known \cite{a3}. Thus, one must check if
it can
be a source for divergences in the entropy in addition to those
generated by $W_{div}({\tilde {\cal M}}_\beta)$.
Fortunately, this question can be resolved because
in the four-dimensional theory the total structure of the surface
divergent terms $W_{div}^{exact}({\cal M}_\beta)$ has been found out
explicitly
\cite{a4}. For the action (\ref{effact}) on the singular space
$W_{div}^{exact}({\cal M}_\beta)$ is written in the
form of (\ref{vol}) where the heat coefficients $a_i({\cal M}_\beta)$
are the sum of the bulk expressions (\ref{a1}), (\ref{a2}) with the
curvature
tensor evaluated in the regular domain of ${\cal M}_{\beta}$ and of
the
additions in the form of the integrals over the horizon surface
$\Sigma$.
For $a_1$
and $a_2$ coefficients these integrals look as
\cite{a4}
\begin{equation}
a_{\beta,1}={\pi\over 3\alpha}(1-\alpha^2)\int_{\Sigma}~~~,
\label{eq:coeff''}
\end{equation}
\begin{equation}
a_{\beta,2}=
{\pi \over 3\alpha}(1-\alpha^2) \int_{\Sigma}\left[
\frac 16 R + {1+\alpha^2 \over 60 \alpha^2}
(2R_{\mu\nu\lambda\rho}n^{\mu}_in^{\lambda}_in^{\nu}_jn^{\rho}_j-
R_{\mu\nu}n^{\mu}_i n^{\nu}_i) \right]~~~.
\label{eq:coeff}
\end{equation}
Remarkably, applying formulas
(\ref{1})-(\ref{4}) one can see that the difference between the heat
coefficients on the regularized space  (when regularization is taken
off)
and on the singular one turns out to be
proportional to the square of the angle deficit at the conical
singularity
$a_i({\tilde {\cal M} }_\beta \rightarrow {\cal M}_\beta)
-a_i({\cal M}_\beta)\sim (1-\alpha)^2$.
This enables one to get the crucial relation
at $\beta\simeq\beta_H$ between the divergences on the regularized
and
singular spaces
\begin{equation}
\lim_{{\tilde {\cal M}}_{\beta} \rightarrow{\cal M}_{\beta}}
W_{div}({{\tilde {\cal M}}_{\beta})=
W_{div}^{exact}({\cal M}_{\beta}}) + O((1-\alpha)^2)~~~.
\label{difference}
\end{equation}
The last term in the right-hand side of (\ref{difference}) comes from
the finite
part $W_{fin}({\tilde{\cal M}}_\beta)$
in the limit ${\tilde {\cal M}}_{\beta} \rightarrow{\cal M}_{\beta}$.
This shows that $W_{fin}({\tilde {\cal M}}_\beta)$ does not influence
the surface divergences  of the order $(1-\alpha)$ and hence they are
completely
removed by the standard renormalization of the gravitational
constants (\ref{const}).
(However, to get rid of the divergent terms of higher order in
$(1-\alpha)$ one is forced to introduce the surface counterterms
additional
to those we have in the
regular case \cite{F}.)

The consequence of equation (\ref{difference}) is that the entropy
at the Hawking temperature
does not acquire additional divergences apart from the standard ones
removed by renormalization of $G$ and $c_i$.
Indeed, from (\ref{difference}) for the divergent part of $S$ one has
$$
S_{div}(\epsilon)
\equiv\left(\beta {\partial \over \partial \beta}-1\right)
W_{div}^{exact}({\cal M}_
{\beta})(\epsilon)|_{\beta=\beta_H}=
$$
\begin{equation}
\left(\beta {\partial \over \partial \beta}-1\right)
\lim_{{\tilde {\cal M}}_{\beta} \rightarrow{\cal M}_{\beta}}
W_{div}({\tilde {\cal M}}_{\beta})(\epsilon)|_{\beta=\beta_H}~~~.
\label{renen}
\end{equation}
Finally, by taking into account (\ref{locentropy}), (\ref{renen}) and
(\ref{renW})
the renormalization of the entropy can be presented as follows:
\begin{equation}
S(G,c_i)+S_{div}(\epsilon)=S(G^{ren},c_i^{ren})~~~.
\label{proof}
\end{equation}
Here $S(G^{ren},c_i^{ren})$ has the form (\ref{locentropy}) expressed
through
$G^{ren}$ and $c_i^{ren}$ related with the bare constants
by the usual equations originated from the one-loop
renormalization (\ref{renW}), (\ref{const}) in quantum theory on
space-times without
horizons.

Equation (\ref{proof}) proves the main statement of this Letter.
It follows from (\ref{proof}) that the finite observed
entropy of a hole $S(G^{ren}, c_i^{ren})$ always comes out as a
combination
of the tree-level bare entropy $S(G, c_i)$ and $S_{div}(\epsilon)$
interpreted as quantum "entanglement" entropy. Thus, if the
gravitational
action is totally induced by quantum effects, then $S(G^{ren},
c_i^{ren})$
is "purebred entanglement entropy" \cite{Jacobson}. Besides, we see
from
(\ref{proof}) that the dependence of $S_{div}(\epsilon)$ on the
number of field species is absorbed into the observable constants
$G^{ren}$
and $c_i^{ren}$.

For simplicity we derived (\ref{proof}) for the scalar model
(\ref{effact}),
but the effect of higher spins can also be incorporated
in our analysis. A special treatment, however, is needed
for the case of nonzero curvature coupling $\xi R\phi^2$
in the scalar Lagrangian \cite{SS}. It should also be noted that our
result concerns the
static black holes and the extension to the stationary geometries is
of interest
as well.

To conclude, the following remarks are in order.
The statistical approach employed here (see Eq. (\ref{entropy}))
gives the
definition of the entropy
which is essentially off-shell.
On the other hand, in the thermodynamical approach one compares
one-loop free energies $F$ of two black hole configurations being in
equilibrium at slightly different temperatures. Since the equilibrium
temperatures take the Hawking values no conical singularities appear
and only the usual (volume) ultraviolet divergences occur in $F$
which are
removed
by the standard renormalization of the gravitational couplings.
This gives the black hole entropy expressed
through the gravitational couplings renormalized from  the very
beginning.
The fact that the statistical and thermodynamical black hole
entropies
are renormalized by the same procedure
seems to be encouraging. Indeed, for usual statistical systems
both the approaches, statistical and thermodynamical, lead to  equal
answers for the entropy. From this point of view the result of this
Letter can be considered as an argument in favor of that the same
correspondence takes place in the black hole physics,
and the Bekenstein-Hawking entropy could have
statistical origin.
However, this problem needs a further investigation.

The scope of this Letter was restricted by the
divergent corrections to the entropy.
However, it is worth pointing out that also finite quantum
corrections
to $S$, that
result in its deviation from the tree-level form (\ref{locentropy}),
are of great interest. Some information about these can be extracted
from the
two-dimensional models where one observes the terms logarithmically
depending on the mass of the hole \cite{a3}. The analogous
terms in four dimensions might be important for understanding the
thermodynamics of quantum holes.

\bigskip\bigskip

This work is partially supported by the International
Science Foundation, grant RFL300, and by the Russian Foundation for
Fundamental Science, grant N 94-04-03665-a.

\bigskip

{\bf Note added} When this paper was in press two new works on the
same subject appeared. In \cite{DLM}  the renormalizability of the
entropy has been demonstrated in the framework of the Pauli-Villars
regularization, whereas in \cite{LW} attention has been paid to
renormalization
of the "area" divergences in the case of vector and spin 1/2 fields.

\end{document}